\documentclass[aps,prd,nofootinbib,amsmath,amssymb,twocolumn,superscriptaddress,10pt]{revtex4}

\pdfoutput=1

\usepackage{graphicx}
\usepackage{epstopdf}
\usepackage{latexsym}
\usepackage{amssymb}
\usepackage{amsmath}
\usepackage{color}
\usepackage{mathrsfs}
\usepackage[center]{subfigure}
\usepackage[colorlinks=true, linkcolor=red, citecolor=blue]{hyperref}

\begin{document}

\newcommand{\bq}{\begin{equation}}
\newcommand{\eq}{\end{equation}}
\newcommand{\bqn}{\begin{eqnarray}}
\newcommand{\eqn}{\end{eqnarray}}
\newcommand{\nb}{\nonumber}
\newcommand{\lb}{\label}
\newcommand{\PRL}{Phys. Rev. Lett.}
\newcommand{\PL}{Phys. Lett.}
\newcommand{\PR}{Phys. Rev.}
\newcommand{\PRD}{Phys. Rev. D}
\newcommand{\CQG}{Class. Quantum Grav.}
\newcommand{\JCAP}{J. Cosmol. Astropart. Phys.}
\newcommand{\JHEP}{J. High. Energy. Phys.}
\newcommand{\PLB}{Phys. Lett. B}

\title{Prospect for cosmological parameter estimation using future Hubble parameter measurements}

\author{Jia-Jia Geng}
\affiliation{Institute  for Advanced Physics $\&$ Mathematics, Zhejiang University of Technology, Hangzhou 310032,  China}
\author{Rui-Yun Guo}
\affiliation{Department of Physics, College of Sciences, Northeastern University, Shenyang 110004, China}
\author{Anzhong Wang}
\affiliation{Institute  for Advanced Physics $\&$ Mathematics, Zhejiang University of Technology, Hangzhou 310032,  China}
\affiliation{GCAP-CASPER, Department of Physics, Baylor University, Waco, TX, 76798-7316, USA}
\author{Jing-Fei Zhang}
\affiliation{Department of Physics, College of Sciences, Northeastern University, Shenyang 110004, China}
\author{Xin Zhang}
\affiliation{Department of Physics, College of Sciences, Northeastern
University, Shenyang 110004, China}
\affiliation{Center for High Energy Physics, Peking University, Beijing 100080, China}
\affiliation{Center for Gravitation and Cosmology, Yangzhou University, Yangzhou 225009, China}


\begin{abstract}

We constrain cosmological parameters using only Hubble parameter data and quantify the impact of future Hubble parameter measurements on parameter estimation for the most typical dark energy models.
We first constrain cosmological parameters using 52 current Hubble parameter data including the Hubble constant measurement from the Hubble Space Telescope.
Then we simulate the baryon acoustic oscillation signals from WFIRST (Wide-Field Infrared Survey Telescope) covering the redshift range of $z\in [0.5, 2]$ and  the redshift drift data from E-ELT (European Extremely Large Telescope) in the redshift range of $z\in [2,5]$.
It is shown that solely using the current Hubble parameter data could give fairly good constraints on cosmological parameters. Compared to the current Hubble parameter data, with the WFIRST observation the $H(z)$ constraints on dark energy would be improved slightly, while with the E-ELT observation the $H(z)$ constraints on dark energy is enormously improved.

\end{abstract}

\pacs{...}

\maketitle

\section{Introduction}
\renewcommand{\theequation}{1.\arabic{equation}} \setcounter{equation}{0}

In exploration of the nature of dark energy, a primary task is to measure the evolution of the equation-of-state (EoS) parameter of dark energy using astronomical observations, based on which theoretical explorations of dark energy would have a clearer direction. However, the measurement of EoS $w(z)$ of dark energy is extremely difficult, owing to the fact that $w(z)$ is actually not an observable, which affects the evolution of the universe, including the expansion history and the growth of structure, in a subtle way. Thus, $w(z)$ of dark energy can only be indirectly measured in the light of effects of dark energy on the evolution of the universe.

Usually, the constraints on $w(z)$ of dark energy are provided by the distance-redshift relation measurements that record the expansion history of the universe, but the cosmic distances, including both luminosity distances from the observations of type Ia supernovae and the angular diameter distances from the observations of baryon acoustic oscillations (BAO), are linked to the EoS of dark energy by an integral over $1/H(z)$, with $H(z)$ being the Hubble parameter of the universe, and $H(z)$ is affected by dark energy via another integral over $w(z)$. Hence, using the distance-redshift relation measurements to constrain $w(z)$ is rather difficult, but using the $H(z)$ measurements to constrain dark energy would become much simpler and more efficient. Obviously, for constraining the EoS of dark energy, the direct measurements of the Hubble parameter at different redshifts are vitally important. Although directly measuring $H(z)$ has been a challenging mission in cosmology, in recent years some $H(z)$ data have been accumulated under the great efforts of astronomers \cite{ZHANG:2012MP,STERN:2009EP,MORESCO:2012JH,Gaztanaga:2008xz,Oka:2013cba,Wang:2016wjr,CHUANG:2012QT,Alam:2016hwk,Moresco:2016mzx,BLAKE:2012PJ,Ratsimbazafy:2017vga,Anderson:2013zyy,Moresco:2015cya,Bautista:2017zgn,Delubac:2014aqe,Font-Ribera:2013wce,Magana:2017nfs,Anderson:2013oza}. It has been shown that, though both the quantity and quality of the current $H(z)$ data are limited, solely using these current data could provide fairly good constraints on some typical dark energy models~\cite{sl5}. For the extensive studies on the use of the current $H(z)$ data in cosmology, see, e.g., Refs.~\cite{Jimenez:2003iv,Simon:2004tf,Zhang:2010ic,Chen:2011ys,H12,Farooq:2012ju,Farooq:2013hq,Melia:2013hsa,Li:2014yza,Sahni:2014ooa,Chen:2013vea,Ding:2015vpa,Chen:2016uno,Farooq:2016zwm,Zheng:2016jlq,Zhang:2017epd,Guo:2017qjt,Leaf:2017kgt,Guo:2018uic,Zheng:2018sxp}.

In the near future, some planned next-generation dark energy experiments, e.g., WFIRST (Wide-Field Infrared Survey Telescope), Euclid, LSST (Large Synoptic Survey Telescope), etc., will be implemented. Thus, both the quantity and quality of $H(z)$ data would be largely enhanced in the next decades. In this paper, we wish to investigate what extent would be achieved by the future next-generation experiments using the $H(z)$ observations in the exploration of the nature of dark energy.


Future Hubble parameter measurements with different facilities complement each other in redshift coverage.
For example, WFIRST observes the BAO signals in the redshift range of $z\in [0.5, 2]$, while E-ELT's (European Extremely Large Telescope) high-resolution optical spectrograph CODEX (COsmic Dynamics EXperiment) observes the Lyman-$\alpha$ absorption lines of distant quasar systems covering the redshift range of $z\in [2,5]$.
In this study, we will simulate future Hubble parameter data from both WFIRST and E-ELT measurements and quantify their impact on parameter estimation.

According to our previous papers~\cite{sl5,sl1,sl2,sl3,sl4,He:2016rvp}, redshift drift measurements from E-ELT produce degeneracy directions in parameter space that are nearly orthogonal to current combined data from different probes, thus can efficiently break degeneracy and significantly improve cosmological constraints. In this paper, we wish to see if the redshift drift observations can play an important role in improving constraints on dark energy when the Hubble parameter data are solely considered.

The paper is organized as follows. In Sec.~\ref{method}, we describe the current and future $H(z)$ data considered in this work. In Sec.~\ref{results}, we present the constraint results and make some relevant discussions. Conclusion is given in Sec.~\ref{conclusion}.


\section{Methodology}\label{method}

In this study, in order to forecast the prospect for cosmological parameter estimation using the future Hubble parameter measurements, we employ three most typical dark energy models, i.e., the $\Lambda$ cold dark matter ($\Lambda$CDM) model in which dark energy is provided by a cosmological constant $\Lambda$ with $w=-1$, the $w$CDM model in which dark energy has a constant EoS $w=w_0$, and the Chevallier-Polarski-Linder (CPL) model in which the EoS of dark energy is dynamically evolutionary as $w(z)=w_0 + w_az/(1+z)$.

To simulate the future $H(z)$ data, we will first constrain the dark energy models by using the current $H(z)$ data, and then use the corresponding best-fit models as fiducial models to produce the mock future data, by which the data inconsistency would be avoided in combining current and future $H(z)$ data in a cosmological fit. We will use the simulated future $H(z)$ data to forecast what extent would be achieved by the future $H(z)$ observations in exploring the nature of dark energy.




\subsection{Current $H(z)$ data, $z\in [0,2.36]$}
 For current Hubble parameter measurements, we use 31 data points from the differential age (DA) and 20 data points from clustering measurements. The DA method proposed by Jimenez and Loeb~\cite{DA} compares the ages of passively-evolving galaxies with similar metallicity and separated by a small redshift interval.
Another independent $H(z)$ measurement used in this paper is from the clustering of galaxies or quasars. It was firstly proposed in Ref.~\cite{Gaztanaga:2008xz}, using the BAO peak position as a standard ruler in the radial direction. Some of the $H(z)$ data points from clustering measurements may be correlated or biased. However, we ignore this problem because it is not the focus of this paper.

 We also use the direct measurement result of the Hubble constant,
 in the light of the cosmic distance ladder from the HST,
 $H_0=73.24\pm 1.74$ km s$^{-1}$ Mpc$^{-1}$~\cite{Riess:2016jrr}.
 Table~\ref{Hz} gives the $H(z)$ data points used in this paper. 51 of these data are from the compilation of Ref.~\cite{Magana:2017nfs}, while the one with $z=0.57$ is from Ref.~\cite{Anderson:2013zyy}.

\begin{table*}
\setlength\tabcolsep{10pt}
\caption{Current Hubble parameter measurements $H(z)$ (in units of km s$^{-1}$Mpc$^{-1}$) and their errors $\sigma_{H}$ at redshift $z$. The method column shows how $H(z)$ was obtained: DA means differential age method, while clustering is from BAO measurements. }
\centering
\begin{tabular}{lllll}
\hline\hline
$ z $ & $ H(z) $ &  $\sigma_{H} $ & Reference & Method \\
\hline
0 & 73.24 & 1.74 & \cite{Riess:2016jrr} & \\
0.07 & 69 & 19.6 & \cite{ZHANG:2012MP} & DA\\
0.1 & 69 & 12 & \cite{STERN:2009EP} & DA\\
0.12 & 68.6 & 26.2 & \cite{ZHANG:2012MP} & DA\\
0.17 & 83 & 8 & \cite{STERN:2009EP} & DA \\
0.1791 & 75 & 4 & \cite{MORESCO:2012JH} & DA\\
0.1993 & 75 & 5 & \cite{MORESCO:2012JH} & DA\\
0.2& 72.9 & 29.6 & \cite{ZHANG:2012MP} & DA\\
0.24 & 79.69 & 2.65 & \cite{Gaztanaga:2008xz} & Clustering \\
0.27 & 77 & 14 & \cite{STERN:2009EP} & DA\\
0.28 & 88.8 & 36.6 & \cite{ZHANG:2012MP} & DA\\
0.3 & 81.7 & 6.22 & \cite{Oka:2013cba}& Clustering \\
0.31 & 78.17 & 4.74 & \cite{Wang:2016wjr}& Clustering \\
0.35 & 82.7 & 8.4 & \cite{CHUANG:2012QT} & Clustering\\
0.3519 & 83 & 14 & \cite{MORESCO:2012JH} & DA\\
0.36 & 79.93  & 3.39 & \cite{Wang:2016wjr}& Clustering \\
0.38 & 81.5  & 1.9 & \cite{Alam:2016hwk}& Clustering \\
0.3802 & 83  & 13.5  & \cite{Moresco:2016mzx} & DA \\
0.4 & 95 & 17 & \cite{STERN:2009EP} & DA\\
0.4004 & 77 & 10.2 & \cite{Moresco:2016mzx} & DA \\
0.4247 & 87.1 & 11.2 & \cite{Moresco:2016mzx} & DA\\
0.43 & 86.45 & 3.68 & \cite{Gaztanaga:2008xz} & Clustering\\
0.44 & 82.6 & 7.8 & \cite{BLAKE:2012PJ} & Clustering\\
0.4497 & 92.8 & 12.9 & \cite{Moresco:2016mzx} &DA \\
0.47 & 89 & 34 & \cite{Ratsimbazafy:2017vga} &Clustering\\
0.4783 & 80.9 & 9 & \cite{Moresco:2016mzx} &DA \\
0.48 & 97 & 60 & \cite{STERN:2009EP} & DA\\
0.51 & 90.4  & 1.9 & \cite{Alam:2016hwk} & Clustering \\
0.52 & 94.35  & 2.65 & \cite{Wang:2016wjr} & Clustering \\
0.56 & 93.33  & 2.32 & \cite{Wang:2016wjr} & Clustering \\
0.57 & 96.8 & 3.4 & \cite{Anderson:2013zyy} & Clustering\\
0.59 & 98.48  & 3.19 & \cite{Wang:2016wjr}& Clustering \\
0.5929 & 104 & 13 & \cite{MORESCO:2012JH} & DA\\
0.6 & 87.9 & 6.1 & \cite{BLAKE:2012PJ} & Clustering \\
0.61 & 97.3  & 2.1 & \cite{Alam:2016hwk} & Clustering \\
0.64 & 98.82  & 2.99 & \cite{Wang:2016wjr} & Clustering \\
0.6797 & 92 & 8 & \cite{MORESCO:2012JH} & DA\\
0.73 & 97.3 & 7 & \cite{BLAKE:2012PJ} & Clustering\\
0.7812 & 105 & 12 & \cite{MORESCO:2012JH} & DA\\
0.8754 & 125 & 17 & \cite{MORESCO:2012JH} & DA\\
0.88 & 90 & 40 & \cite{STERN:2009EP} & DA\\
0.9 & 117 & 23 & \cite{STERN:2009EP} & DA\\
1.037 & 154 & 20 & \cite{MORESCO:2012JH} & DA\\
1.3 & 168 & 17 & \cite{STERN:2009EP} & DA\\
1.363 & 160 & 33.6 & \cite{Moresco:2015cya} & DA\\
1.43 & 177 & 18 & \cite{STERN:2009EP} & DA\\
1.53 & 140 & 14 & \cite{STERN:2009EP} & DA\\
1.75 & 202 & 40 & \cite{STERN:2009EP} & DA\\
1.965 & 186.5 & 50.4 & \cite{Moresco:2015cya} & DA \\
2.33 & 224 & 8 & \cite{Bautista:2017zgn} & Clustering\\
2.34 & 222 & 7 & \cite{Delubac:2014aqe}& Clustering\\
2.36 & 226 & 8 & \cite{Font-Ribera:2013wce}& Clustering\\
\hline\hline
\end{tabular}
\label{Hz}
\end{table*}

\subsection{Future $H(z)$ data from WFIRST, $z\in [0.5,2]$}

In order to examine future Hubble parameter data, we simulate mock BAO data using the method described in Ref.~\cite{DETF} and consider the future BAO data based on the long-term space-based project WFIRST. For the details, we refer the reader to Ref.~\cite{DETF}. These BAO data are uniformly distributed in 10 redshift bins of $z\in [0.5,~2]$, with each $\Delta z_i$ centered on the grid $z_i$. The observables in Ref.~\cite{DETF} are the expansion rate $H(z)$ and the comoving angular diameter distance $d_A^{co}(z)=d_L(z)/(1+z)$. However, we are only interested in the observable $H(z)$ and ignore the other one in this paper. The uncertainty of $\ln H(z_i)$ can be written as

\begin{equation}\label{eqH}
\sigma_H^i=x_0^H\frac{4}{3}\sqrt{\frac{V_0}{V_i}}f_{nl}(z_i).
\end{equation}
Here $V_i=1500(d_A^{co}(z_i))^2/H(z_i)$ is the comoving survey volume in the redshift bin of $z_i$, while the erasure of the baryon features by non-linear evolution is factored in using $f_{nl}(z_i)=1$ for $z_i>1.4$ and $f_{nl}(z_i)=(1.4/z_i)^{1/2}$ for $z_i<1.4$.
We also consider the systematic errors which are modeled as independent uncertainties in the log of the distance measures in each redshift bin: $\sigma_s^i=0.01\sqrt{\frac{0.5}{\Delta{z_i}}}$, with $\Delta{z_i}=0.15$. In our simulation we choose $x_0^H=0.0148$, $x_0^d=0.0085$, and $V_0=\frac{2.16}{h^3}$ Gpc$^3$.


\subsection{Future redshift drift data from E-ELT, $z\in [2,5]$}

The drift in the redshift of observed objects passively follows the cosmological expansion. Sandage~\cite{sandage} firstly pointed out that one could directly measure the variation of redshift of distant sources. Then Loeb~\cite{loeb} presented the possibility of detecting redshift drift in the spectra of Lyman-$\alpha$ forest of distant quasars (QSO) in decades. The E-ELT equipped with a high-resolution spectrograph called CODEX is to be built to achieve this goal.

The observable in the redshift drift method is
\begin{equation}\label{eq6}
\ \Delta v \equiv \frac{\Delta z}{1+z}=H_0\Delta t_o\bigg[1-\frac{E(z)}{1+z}\bigg],
\end{equation}
where $\Delta t_o$ is the time interval of observation, and $E(z)=H(z)/H_0$ is given by a specific dark energy model.

From Eq.~(\ref{eq6}), we have
$H(z)=(1+z)(H_0-\Delta v/\Delta t_o)$ and $\sigma_{H(z)}=(1+z)\sigma_{\Delta v}/\Delta t_o$.
Therefore, we can simulate future $H(z)$ data using the redshift drift method, which covers the redshift range of $z\in[2,5]$.
The magnitude of the redshift drift is minuscule, i.e., of order several cm s$^{-1}$ at redshift $z\simeq1$.

In a flat universe, we have
\begin{equation}\label{Ez}
E(z)=\sqrt{\Omega_r(1+z)^4+\Omega_m(1+z)^3+(1-\Omega_r-\Omega_m){X}(z)},
\end{equation}
where $\Omega_r$ and $\Omega_m$ are the present-day density parameters of radiation and matter, respectively, and ${X}(z)\equiv\rho_{\rm de}(z)/\rho_{\rm de}(0)=\exp[3\int_0^z {1+w(z')\over 1+z'}dz']$.

According to Ref.~\cite{Liske:2008ph}, the uncertainty of $\Delta v$ measurements expected by E-ELT can be expressed as
\begin{equation}\label{eq7}
\sigma_{\Delta v}=1.35
\bigg(\frac{S/N}{2370}\bigg)^{-1}\bigg(\frac{N_{\mathrm{QSO}}}{30}\bigg)^{-1/2}\bigg(
\frac{1+z_{\mathrm{QSO}}}{5}\bigg)^{x}~\mathrm{cm}~\mathrm{s}^{-1},
\end{equation}
with $x=-1.7$ for $2<z<4$ and $x=-0.9$ for $z>4$.
$S/N = 3000$ is the signal-to-noise ratio defined per 0.0125 ${\AA}$ pixel, $N_{\mathrm{QSO}}$ is the
number of observed quasars, while $z_{\mathrm{QSO}}$ represents their redshift.

To simulate the redshift drift data, we first constrain the dark energy models by using the current Hubble parameter data. The obtained best-fit parameters are substituted into Eq.~(\ref{eq6}) to get the central values of the redshift drift data.
We choose $N_{\mathrm{QSO}}=30$ mock data uniformly distributed among
six redshift bins of $z_{\rm QSO}\in [2, 5]$ and typically take $\Delta t_o=30$ yr in our analysis.
The error bars are computed from Eq.~(\ref{eq7}).


\section{Results and Discussion}\label{results}

\begin{figure}
\begin{center}
\includegraphics[width=9cm]{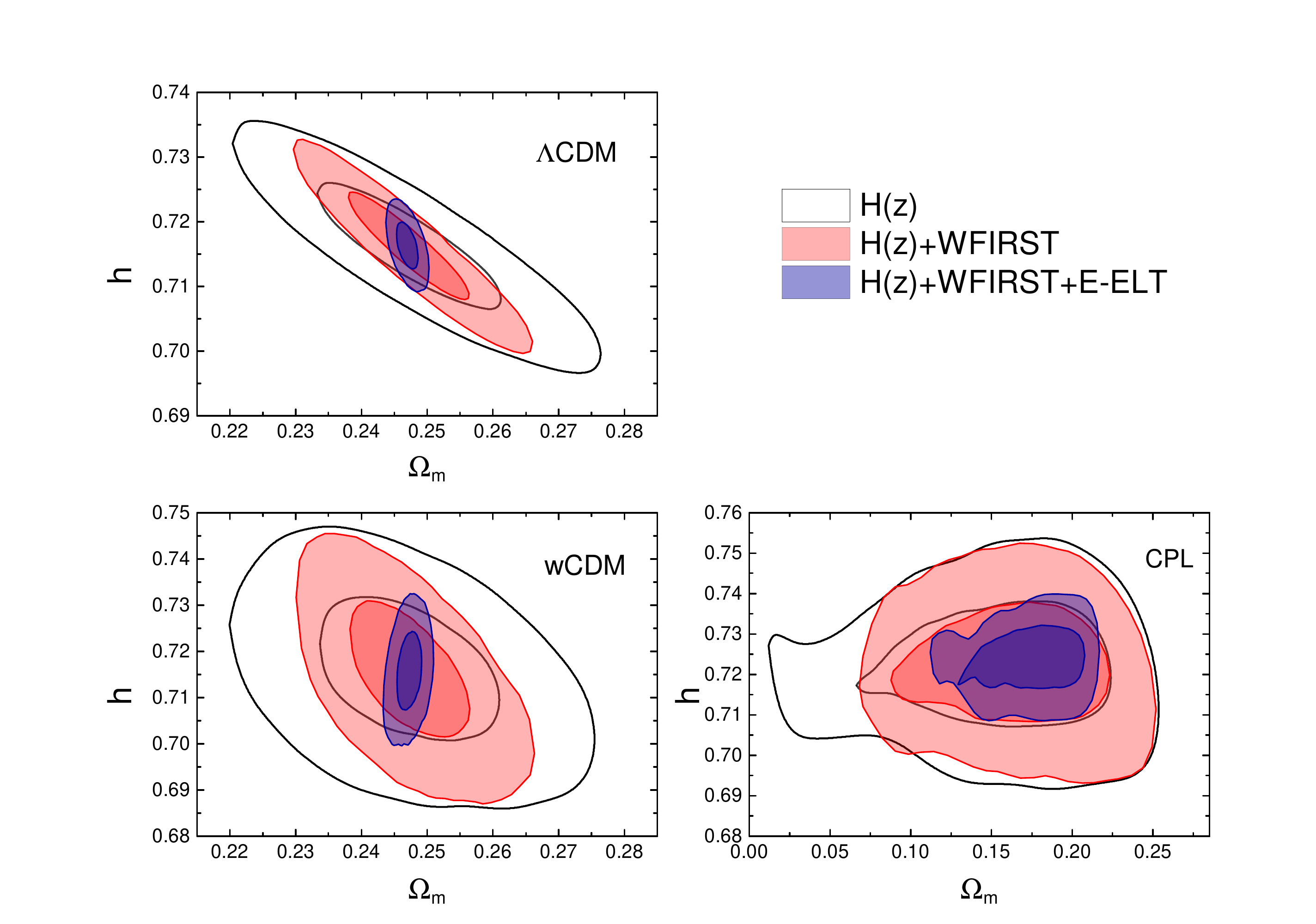}
\end{center}
\caption{Constraints (68.3\% and 95.4\% CL) in the $\Omega_m$--$h$ plane for $\Lambda$CDM, $w$CDM, and CPL models with the $H(z)$, $H(z)$+WFIRST, and $H(z)$+WFIRST+E-ELT data.}
\label{fig1}
\end{figure}

\begin{table*}\tiny
\caption{Errors of parameters in the $\Lambda$CDM, $w$CDM, and CPL models for the fits to
the $H(z)$, $H(z)$+WFIRST, and $H(z)$+WFIRST+E-ELT data.}
\label{table2}
\small
\setlength\tabcolsep{2.8pt}
\renewcommand{\arraystretch}{1.5}
\centering
\begin{tabular}{cccccccccccc}
\\
\hline\hline &\multicolumn{3}{c}{$H(z)$} &&\multicolumn{3}{c}{$H(z)$+WFIRST} &&\multicolumn{3}{c}{$H(z)$+WFIRST+E-ELT } \\
           \cline{2-4}\cline{6-8}\cline{10-12}
Error  & $\Lambda$CDM & $w$CDM & CPL & & $\Lambda$CDM & $w$CDM & CPL & & $\Lambda$CDM & $w$CDM & CPL\\ \hline

$\sigma(w_0)$              & $-$
                   & $0.0802$
                   & $0.1561$&
                   & $-$
                   & $0.0684$
                   & $0.1310$&
                   & $-$
                   & $0.0498$
                   & $0.0904$\\

$\sigma(w_a)$              & $-$
                   & $-$
                   & 0.3944&
                   & $-$
                   & $-$
                   & $0.3797$&
                   & $-$
                   & $-$
                   & $0.2467$
                   \\

$\sigma(\Omega_{m})$       & $0.0146$
                   & $0.0144$
                   & $0.0897$&
                   & $0.0094$
                   & $0.0094$
                   & $0.0724$&
                   & $0.0017$
                   & $0.0020$
                   & $0.0407$\\

$\sigma(h)$              & $0.0098$
                   & $0.0157$
                   & $0.0160$&
                   & $0.0084$
                   & $0.0148$
                   & $0.0152$&
                   & $0.0036$
                   & $0.0086$
                   & $0.0081$\\

\hline
\hline
\end{tabular}
\end{table*}

Fig.~\ref{fig1} shows the joint constraints on the $\Lambda$CDM, $w$CDM, and CPL models in the $\Omega_m$--$h$ plane. Here, $\Omega_m$ is the present-day matter density, and $h=H_0/(100~{\rm km~s~Mpc^{-1}})$.
The 68.3\% and 95.4\% CL posterior distribution contours are shown. The data combinations used are the $H(z)$ data, the $H(z)$+WFIRST data, and the $H(z)$+WFIRST+E-ELT data, and their constraint results are shown with white, red, and blue contours, respectively.
The 1$\sigma$ errors of the parameters $w_0$, $w_a$, $\Omega_m$, and $h$ for the three models for the above three data combinations are given in Table~\ref{table2}.
It is shown that the current $H(z)$ data could provide fairly good constraints on these typical dark energy models.
With the $H(z)$+WFIRST observation, the constraints on $\Omega_m$ and $h$ will be improved, respectively, by
35.6\% and 14.3\% for the $\Lambda$CDM model, by 34.7\% and 5.7\% for the $w$CDM model, and by $19.3\%$ and 5.0\% for the CPL model.
With the $H(z)$+WFIRST+E-ELT observation, the constraints on $\Omega_m$ and $h$ will be further improved by
81.9\% and 57.1\% for the $\Lambda$CDM model, by 78.7\% and 41.9\% for the $w$CDM model, and by $43.8\%$ and 46.7\% for the CPL model, respectively, compared to those with the $H(z)$+WFIRST observation. This is because the degeneracy between $\Omega_m$ and $h$ can be well broken with the E-ELT data, which is shown in Fig.~\ref{fig1}.
Therefore, we conclude that with the WFIRST observation the $H(z)$ constraints on dark energy would be improved slightly, while with the E-ELT observation the $H(z)$ constraints on dark energy would be enormously improved.

\begin{figure}
\begin{center}
\includegraphics[width=9cm]{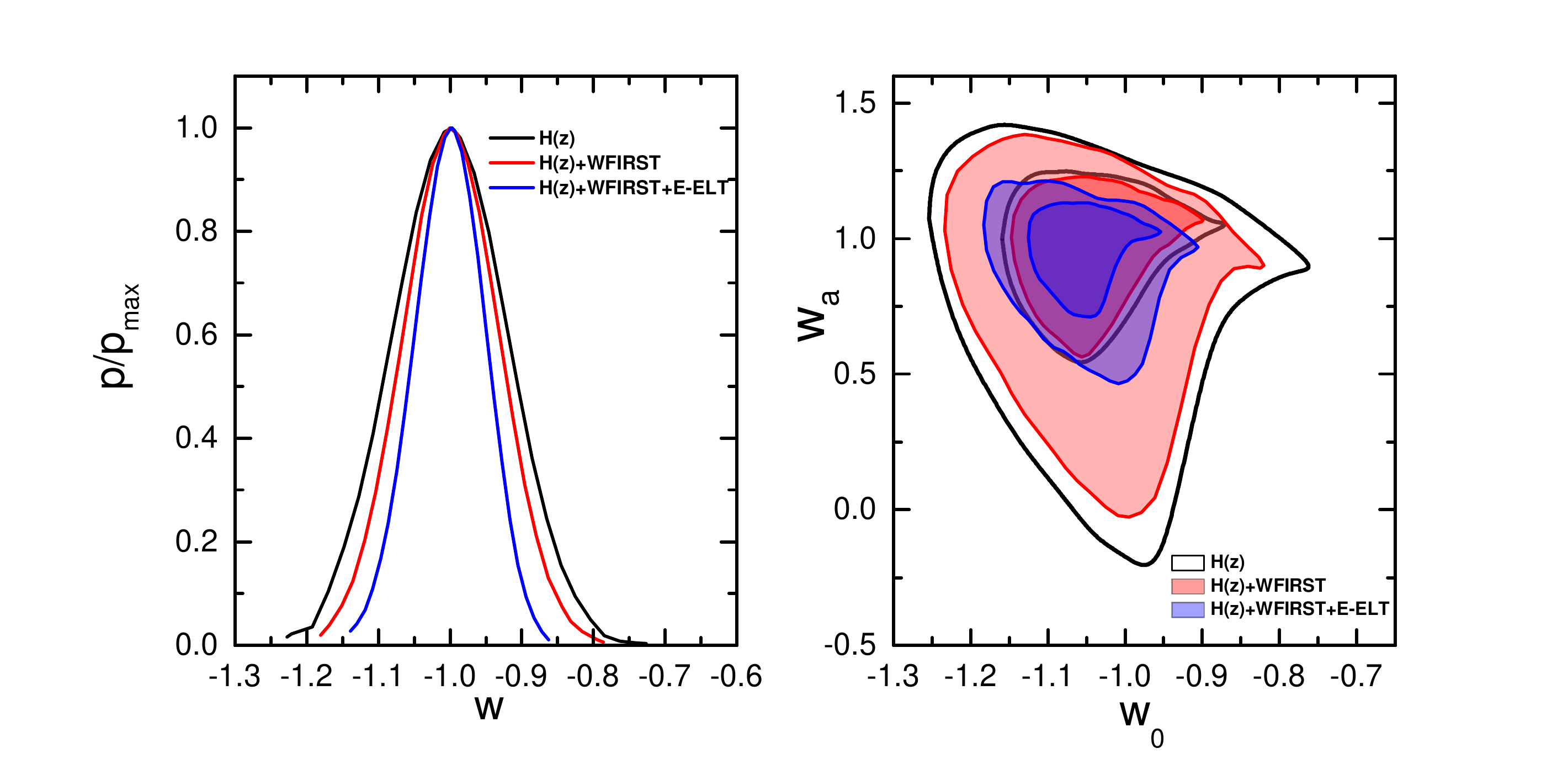}
\end{center}
\caption{The one-dimensional posterior distributions of $w$ for the $w$CDM model (left) and the two-dimensional posterior distributions of $w_0$ and $w_a$ for the CPL model (right), from the $H(z)$, $H(z)$+WFIRST, and $H(z)$+WFIRST+E-ELT  constraints.}
\label{fig2}
\end{figure}

We also discuss the impact of future Hubble parameter measurements on constraining the EoS of dark energy.
In Fig.~\ref{fig2} we show the one-dimensional posterior distributions of $w$ for the $w$CDM model and the two-dimensional posterior distributions of $w_0$ and $w_a$ for the
CPL model, from the $H(z)$, $H(z)$+WFIRST, and $H(z)$+WFIRST+E-ELT constraints.
The corresponding errors of $w_0$ and $w_a$ are given in Table~\ref{table2}.
With the $H(z)$+WFIRST observation,
the constraint on $w$ will be improved by 14.7\% for the $w$CDM model, while the constraints on $w_0$ and $w_a$ will be improved by $16.1\%$ and 3.7\% for the CPL model, respectively.
With the $H(z)$+WFIRST+E-ELT observation,
the constraint on $w$ will be further improved, by 27.2\% for the $w$CDM model, while the constraints on $w_0$ and $w_a$ will be further improved by $31.0\%$ and 35.0\% for the CPL model, compared to those with the $H(z)$+WFIRST observation.
Therefore, with the WFIRST observation the constraints on the EoS of dark energy would be improved slightly, while with the E-ELT observation the constraints on the EoS of dark energy would be further enormously improved.

\section{Conclusion}\label{conclusion}

In this paper, we forecast the prospect for constraining dark energy models using future Hubble parameter measurements. We show that direct measurements of the Hubble parameter at different redshifts are rather important for cosmological parameter estimations.
Solely using the current $H(z)$ data could provide fairly good constraints on typical dark energy models.
We quantify the impact of future Hubble parameter observations on cosmological parameter estimations.
In order to avoid data inconsistency, the best-fit models based on current data are chosen as the fiducial models to simulate mock future $H(z)$ data.
We simulate future Hubble parameter data from the WFIRST and E-ELT observations, which cover the redshift ranges of $z\in [0.5, 2]$ and $z\in [2, 5]$, respectively.
 It is shown that with the $H(z)$+WFIRST observation, the parameter estimation results will be improved slightly.
 Furthermore, we find that redshift drift data from E-ELT can effectively break the degeneracy between $\Omega_m$ and $h$. Therefore, with the $H(z)$+WFIRST+E-ELT observation, the constraints on $\Omega_m$ and $h$ will be greatly improved.
 We also discuss the impact of future $H(z)$ measurements on constraining the EoS of dark energy and show that with the E-ELT observation the constraints on the EoS of dark energy would be enormously improved.

\acknowledgments
This work was supported by the National Natural Science Foundation of China (Grants No.~11522540, No.~11690021, No.~11375153 and No.~11675145), the National Program for Support of Top-Notch Young Professionals, and the 2016 Program for Postdoctoral Fellowship of Zhejiang Province.

\end{document}